\documentclass{aa} 
\usepackage{graphicx,subfigure}
\begin{document}     

\title{Luminosity function of clusters of galaxies}     

\author{M. Paolillo\inst{1}
          \fnmsep\thanks{
		Universit\`a di Palermo, D.S.F.A., Sez.di Astronomia
              P.zza del Parlamento 1, 90134 Palermo, Italy;
	      email: paolillo@astropa.unipa.it}
	      \and S. Andreon\inst{1}\and G. Longo
   \inst{1}\and E. Puddu\inst{1} \and R. R. Gal\inst{3}  
     \and R. Scaramella \inst{2} \and S. G. Djorgovski\inst{3}
     \and R. de Carvalho\inst{4}     
              }                
   \offprints{G. Longo, \email{longo@na.astro.it}}
     
   \institute{
	      Osservatorio Astronomico di Capodimonte, via Moiariello 16,
	      80131 Napoli, Italy
	    \and
	      Osservatorio Astronomico di Monte Porzio, via Frascati 33,
	      00044 Roma, Italy
	    \and
	      Department of Astronomy, Caltech, USA
	    \and
	      Observat\'orio Nacional, Rua General Jos\'e Cristino 77, 20921
	      - 400 Rio de Janeiro, Brazil  
	      }
\date{Received 22 September 2000/ Accepted 5 December 2000}     
\abstract{ 
The composite galaxy luminosity function (hereafter LF) of 39 Abell clusters of
galaxies is derived by computing the statistical excess of galaxy counts in the
cluster direction with respect to control fields.
Due to the wide field
coverage of the digitised  POSS-II plates, we can measure field counts around
each cluster in a fully homogeneous way.
Furthermore, the availability of virtually unlimited sky coverage allows us to 
directly compute the LF errors without having to rely 
on the estimated variance of the background.
The wide field coverage also allows us to derive the LF of the whole
cluster, including galaxies located in the cluster outskirts.
The global composite LF has a slope
$\alpha\sim-1.1\pm0.2$ with minor variations from blue to red filters, and
$M^*\sim-21.7,-22.2,-22.4$ mag ($H_0=50$ km s$^{-1}$ Mpc$^{-1}$) in $g, r$ and $i$
filters, respectively (errors are detailed in the text). These results are in quite good
agreement with several previous determinations and in particular
with the LF determined for the inner region of a largely overlapping set of
clusters, but derived making use of a completely different method for background subtraction. The
similarity of the two LFs suggests the existence of minor differences between the LF in the cluster
outskirts and in the central region, or a negligible
contribution of galaxies in the cluster outskirts to the global LF. 
}  
\maketitle     
\section{Introduction} 

The galaxy luminosity function (hereafter LF) measures the relative frequency of
galaxies as a function of luminosity per unit
co-moving volume. Thus, the LF is the zero--order statistic of galaxy
samples and provides the natural weighting of most statistical quantities.  For instance,
the luminosity evolution is often inferred by the variation with  redshift of the LF; 
the metal production rate is the integral of the luminosity weighted against the LF; 
the fraction of blue galaxies, crucial for the Butcher--Oemler effect  (Butcher 
\& Oemler \cite{Butch84}), is given by the ratio between the color distribution,
averaged over the LF, and the total
number of galaxies (i.e. the integral of the LF). 
The LF is, therefore, central to many cosmological issues (Binggeli, Sandage \& Tammann \cite{Bin88};
Koo \& Kron \cite{Koo92}; Ostriker \cite{Ostr93}).

The determination of the {\it cluster} LF is observationally less expensive than the
analogous determination of the {\it field} LF. In fact, the cluster LF can be
determined as the statistical excess of galaxies along the cluster line of sight, with
respect to the control field direction, due to the fact that clusters appear as overdensities with 
respect to the intracluster field. Therefore we do not need to know the redshift of each cluster member but only the mean cluster redshift, provided that we treat the sampled volume as a free parameter.
This approach assumes implicitly that the background contribution
along the cluster line of sight is
equal to the ``average'' background, a hypothesis that a non-zero correlation
function for galaxies shows to be only approximate: there are galaxies near
the cluster line of sight, but not belonging to the cluster itself, in excess of the value 
expected by assuming a uniform ``average'' galaxy density. In other words,
it happens very often that a nearby group, cluster or supercluster
contaminates the control field counts or the cluster counts 
thus affecting the determination of the cluster LF. 
This problem is even more relevant when sampling the cluster outskirts, where galaxy evolution
probably occurs (van Dokkum et al. \cite{vanD98}) since i) the low galaxy density  
of these regions is affected by even a few contaminants, and
ii) the large observing area makes more probable the presence  of a contaminating group.
Recently, Huang et al. (\cite{Huang97}) found an expression for estimating the error
introduced by a non zero correlation function. This expression however, inflates
errors as a consequence of the fact that the statistics is not simply
Poissonian, and does not try to correct field counts to the value expected once the contribution due to other prospectically near  
overdensities is taken into account.

From an observational point of view, a proper determination of the LF 
with small field of view imagers and in presence of a non--zero correlation
function is very time consuming since several fields
all around the cluster need to be observed to estimate the
field counts along the clusters line of sight. Therefore, in order to save precious
telescope time, very
often the field counts are taken from the literature (and usually concern a
specific region of the sky which is often completely unrelated to the cluster line of sight)
or only a few (usually one, except Bernstein et al. \cite{Ber95}) 
comparison fields at fairly different right ascensions are adopted. The alternative route
is to recognize cluster membership individually, for instance on
morphological grounds as Binggeli, Sandage
\& Tammann (\cite{Bin85}) did for the Virgo cluster, 
or by means of galaxy colors, as in Garilli, Maccagni \& Andreon (\cite{Garilli99}, hereafter GMA99).

Wide--field imagers, such as Schmidt plates or large CCD mosaics, allow one instead to
sample lines of sight all around the cluster, and
accurately determine the field properties along the cluster line of
sight (cf. Valotto et al. \cite{Val97}).

Our group is currently exploiting the Digitized Palomar Sky Survey
(DPOSS) and the resulting Palomar-Norris Sky Catalog (PNSC) in the context of the CRoNaRio collaboration
({\bf C}altech--{\bf Ro}ma--{\bf Na}poli--{\bf Rio})(Djorgovski et
al. \cite{Djor99}). 
Due to the good photometric quality of the data and the wide sky coverage of DPOSS data, the survey is
particularly tailored to explore
the actual background contribution to the determination of the cluster
LF.\\

This paper is organised as follow:
in Sec. 2 we briefly describe the main characteristics of the data and we
present the cluster sample. Sec. 3 deals with most technical problems related to
the determination of the individual LF of clusters. Sec. 4 presents the results of
this work and a comparison with literature results. Conclusions are summarised in
Sec.5.
We adopt $H_0=50$ km s$^{-1}$ Mpc$^{-1}$ and $q_0=0.5$.


   \begin{table}[htb]
   \caption[]{The cluster sample}
   \begin{tabular}{ccccc}
\hline
\\
{Cluster} & {Redshift} & {Plate} & {Richness class} & {B-M type}\\
\\
\hline
\\
A1 & 0.1249 & 607 & 1 & III\\
A16 & 0.0838 & 752 & 2 & III\\
A28 & 0.1845 & 680 & 2 & III\\
A41 & 0.2750 & 752 & 3 & II-III\\
A44 & 0.0599 & 680 & 1 & II\\
A104 & 0.0822 & 474 & 1 & II-III\\
A115 & 0.1971 & 474 & 3 & III\\
A125 & 0.188 & 610 & 1 & III\\
A150 & 0.0596 & 610 & 1 & I-II \\
A152 & 0.0581 & 610 & 0 & ...\\
A158 & 0.0645 & 610 & 0 & ...\\
A180 & 0.1350 & 755 & 0 & I\\
A192 & 0.1215 & 755 & 2 & I\\
A202 & 0.1500 & 755 & 2 & II-III\\
A267 & 0.2300 & 829 & 0 & ... \\
A279 & 0.0797 & 829 & 1 & I-II\\
A286 & 0.1603 & 829 & 2 & II\\
A293 & 0.1650 & 757 & 2 & II\\
A294 & 0.0783 & 757 & 1 & I-II\\
A1632 & 0.1962 & 443 & 2 & II-III\\
A1661 & 0.1671 & 443 & 2 & III\\
A1677 & 0.1845 & 443 & 2 & III\\
A1679 & 0.1699 & 443 & 2 & III\\
A1809 & 0.0788 & 793 & 1 & II\\
A1835 & 0.2523 & 793 & 0 & ...\\
A2049 & 0.1170 & 449 & 1 & III\\
A2059 & 0.1305 & 449 & 1 & III\\
A2061 & 0.0782 & 449 & 1 & III\\
A2062 & 0.1122 & 449 & 1 & III\\
A2065 & 0.0721 & 449 & 2 & III\\
A2069 & 0.116 & 449 & 2 & II-III\\
A2073 & 0.1717 & 449 & 1 & III\\
A2083 & 0.1143 & 449 & 1 & III\\
A2089 & 0.0743 & 449 & 1 & II\\
A2092 & 0.066 & 449 & 1 & II-III\\
A2177 & 0.1610 & 517 & 0 & ...\\
A2178 & 0.0928 & 517 & 1 & II\\
A2223 & 0.1027 & 517 & 0 & III\\
A2703 & 0.1144 & 607 & 0 & ...\\
\\
\hline
	\end{tabular}
   \label{sampletab}
\end{table}

   \begin{figure*}
   \sidecaption
   \includegraphics[width=12cm]{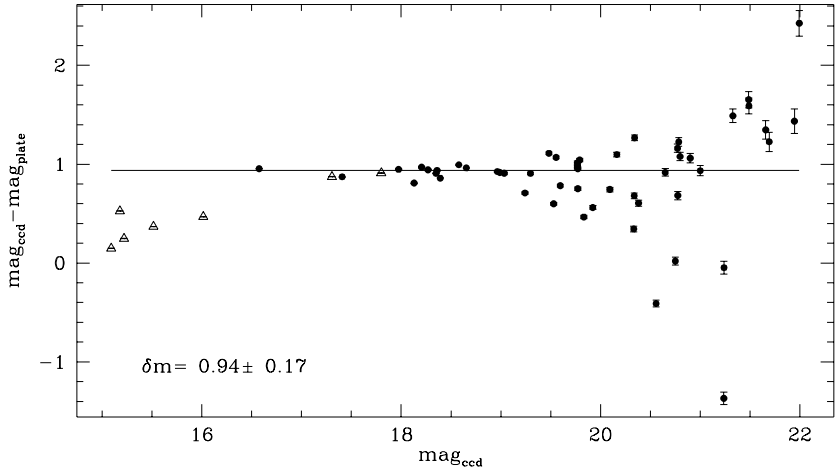}
      \caption{Comparison between aperture instrumental magnitudes
	measured on the photographic plate and CCD aperture  
	magnitudes. The continuous line represents the
	median difference $\delta
	m=mag_{ccd}-mag_{plate}$ for galaxies (filled
	dots). Stars (empty triangles) are excluded from the fit because
	they are often saturated on POSS II plates. This diagram has been
	used for calibrating the plate $F449$.}
         \label{cal}
   \end{figure*}  

\section{The data \& the sample}

The data used in this paper were extracted from the DPOSS
frames taken in the photographic $J, F$ and $N$ bands (Reid et al.
\cite{Reid91}). Weir et al.(\cite{Weir95c})  describe the 
characteristics of the SKICAT package, which performs the plate linearization
and the object detection and classification (based on a classification tree,
see Weir, Fayyad \& Djorgovski \cite{Weir95a}).

SKICAT measures four different magnitudes for each object 
detected on the plates, among which the 
FOCAS (Jarvis \& Tyson \cite{Jarvis81}) total magnitude, obtained by dilatation
of the detection isophote in all directions until the object area is doubled.
These magnitudes approximate true asymptotic magnitudes.

The plates are individually calibrated to the Gunn system
(Thuan \& Gunn  \cite{Thuan76}; Wade \cite{Wade79}) by means of CCD frames of
clusters of galaxies. We used the data set presented in Garilli et al. 
(\cite{Garilli96}), which has been used in GMA99 to compute the cluster LF. As they point out, their
Gunn $g$ photometry does not perfectly match the standard Thuann--Gunn system (for
historical reasons): $g_{Garilli}=g-0.20\pm 0.14$.  However, the
error is systematic, so that we recover the true Gunn $g$ magnitude by
adding this offset. We note that this is different from the general CCD 
calibration of DPOSS/PNSC, which is mainly  based on the extensive CCD data
 sets obtained at Palomar for this purpose (Gal et al. \cite{Gal2000}).

Plates are photometrically calibrated by comparing plate and CCD aperture
(within 5 arcsec radius) photometry of common galaxies, and magnitudes are corrected for
Galactic absorption. A typical calibration diagram is shown in Fig.\ref{cal}.
The adopted zero point is the median of the differences
$mag_{ccd}-mag_{plate}$, after excluding bright stars (empty
triangles) that are usually saturated on photographic plates.
No color term has been adopted as required by the POSS-II photometric
system (Weir, Djorgovski \& Fayyad \cite{Weir95b}).
The mean error\footnote{We adopt as errors on our median
zero point the semi-interquartile intervals.} on the zero-point determination
is 0.02 mag in $g$ and 0.04 mag in $r$ and $i$, while the typical photometric error
on individual magnitudes (including Poissonian errors, residuals of density to intensity
conversion, etc.) is 0.2 mag in $g$ and 0.16 in $r$ and $i$.
$K$-corrections were taken from Fukugita et al. (\cite{Fuk95}). Our data do not
have enough resolution to distinguish between different morphological types, nor this selection can be done using galaxy colors due to the errors on individual magnitudes. Anyway the difference in k-corrections between E and Scd is $\leq 0.25$ mag in the $r$ and $i$ bands for the most distant cluster in our sample 
($\leq 0.3$ mag at our median redshift in all bands) so that we could adopt the k-correction of the dominant E-S0 population.

We estimated the photometric completeness limit of our data for each cluster and in
each band independently, in order to take into account the depth variations of our
catalogs from plate to plate and as a function of the projected cluster location on the
plate. We adopt as our completeness limit the magnitude at which nearby field counts systematically deviate from linearity (in logarithmic units).
The use of homogeneous data, reduced in one single way, both for the
control field and the cluster galaxy counts, helps to partially compensate for
systematic errors due to selection effects which cancel out (at least in part) in
the statistical subtraction of the counts.

The studied sample is extracted from the  Abell catalogue (Abell
\cite{Abell58}; Abell, Corwin \& Olowin \cite{ACO89}), among those clusters with known redshift,
which are imaged in a fully reduced plate triplet (i.e. $J, F$ 
and $N$) and with photometric
zero points already available to us, in all three bands, 
at the start of this work. At that time, 39 Abell clusters
satisfied the above conditions, the bottleneck being due to the low number of
calibration frames and the requirement of having at least one reliable spectroscopic 
redshift for the cluster.\\

A few more clusters satisfying the above
condition were also rejected from the sample on the following grounds:\\
{\it Abell 154} 
- There are two density peaks at two different redshifts, along the line of sight, respectively
at $z=0.0640$ (A154) and $z=0.0428$ (A154a).\\
{\it Abell 156} 
- There are two discordant redshift measurements in the literature.
Since there is no galaxy  
overdensity at the cluster position we can safely assume that it is 
a spurious object.\\
{\it Abell 295} 
- Two density peaks in the cluster direction: A295 at 
$z=0.0424$ and A295b at $z=0.1020$.\\
{\it Abell 1667} 
- Two density peaks in the cluster direction: A1667 at
$z=0.1648$ and A1667b at $z=0.1816$.\\
{\it Abell 2067} 
- Two density peaks in the cluster direction: A2067 at
$z=0.0756$ and A2067b at $z=0.1130$.\\
Two more clusters, Abell 158 and Abell 259, show a double structure with
two adjacent but distinct density peaks. In these cases we included only the
galaxies belonging to the peaks with measured redshift, without assuming 
that the secondary peak lies at the same redshift as the first one.  
The final sample is listed in Table~\ref{sampletab}.

\section{Individual LF determination} 
\subsection{Measure of the background counts and of their variance}

In order to accurately compute the cluster LF,
we need to statistically subtract the background from 
galaxy counts in the cluster direction. This step requires particular
attention to three potential sources of errors:

   \begin{figure}[t]
   \resizebox{\hsize}{!}{\includegraphics{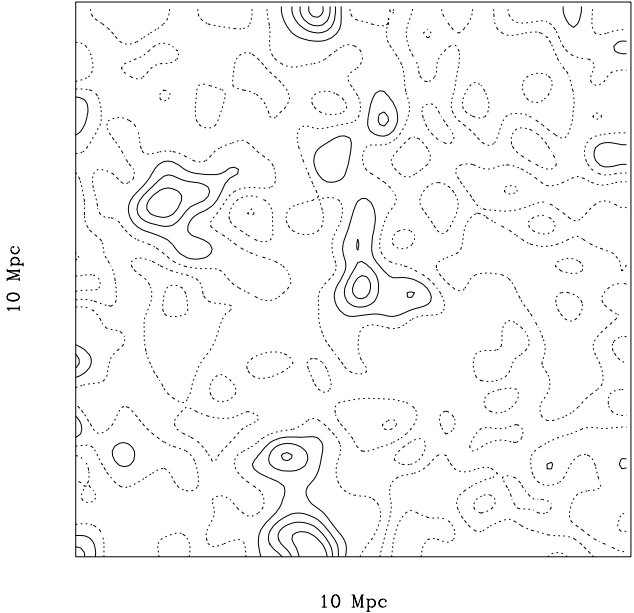}}
      \caption[]{Density map of a $10\times 10$ Mpc region around Abell 152.
       The 
	continuous and dotted lines represent density levels
	respectively above and under the $1.5\sigma$
	level used to detect the cluster area (sec.3.2). Levels are
	spaced $1\sigma$ apart. }
         \label{densmap}    \end{figure}

   \begin{figure}[th]
   \resizebox{\hsize}{!}{\includegraphics{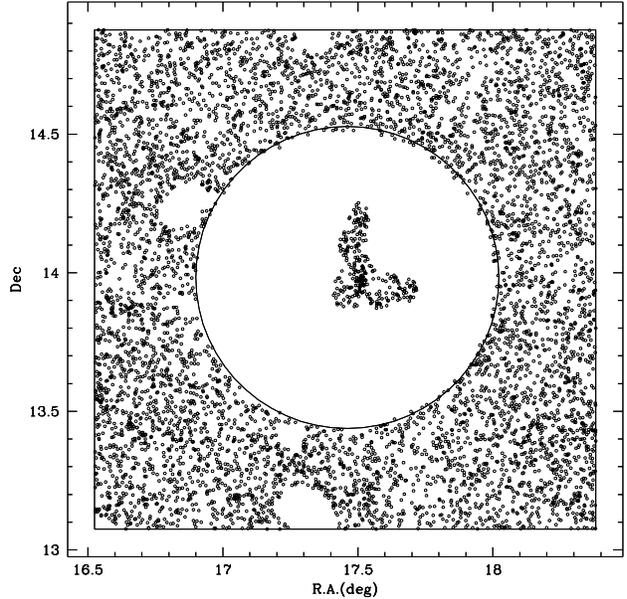}}
      \caption[]{The Abell 152 cluster+background field. Dots
	within the inner circle represent galaxies included in the $1.5\sigma$
	region, while those outside the circle are background galaxies.
	The empty regions in the background area represent the removed
	overdensity regions (sec 3.1). The region is the same as shown in
	Fig.\ref{densmap}.}
         \label{a152}    \end{figure}

1) Catalogs should be complete and clean from spurious detections: 
one single entry in the catalog should
correspond to each object in the sky and viceversa.
Unfortunately, single-filter SKICAT catalogs contain numerous faint spurious
objects, mainly around bright sources. This is due to the fact that SKICAT was
intentionally tailored to 
detect even the faintest objects and therefore pays the price of detecting false objects in the 
halo of saturated or extended sources.
 
All objects in a $10 \times 10$ Mpc region centered on the cluster
center (as estimated by Abell, 1958), are thus extracted from the three calibrated catalogs 
(in $g, r$ and $i$) and positionally matched using their right
ascension and declination.
False detections are removed in the matching step, due
to the low probability that two (or even three) false detections occur in the same
sky position in different filters. 
The maximum
allowed distance for the matching  was fixed at 7 arcsec in order to take into account
the positional uncertainty (2.1
arcsec at 95\% confidence level (Deutsch \cite{Deu99}) for a single filter) while
minimizing the number of erroneous matches (the average distances between galaxies in the cluster regions is 5 times our matching distance).
As a further precaution, we excluded a circular region (with an area five
times larger than the isophotal area) around each potential troublemaker 
(bright stars and very extended objects). The coordinates of the
removed areas were stored to allow for later corrections. The total area removed
constitutes, at most, 3\% of the overall area.

The final catalogue was then used to produce a galaxy surface density map which (in order to enhance structures at the cluster scale, Fig.\ref{densmap})
was convolved with a Gaussian function having $\sigma$ of 250 Kpc -- i.e. the
size of a typical core radius -- in the cluster rest frame.

2) Counts in the cluster and control field directions should be accurately
photometrically calibrated without relative systematic errors.
We achieve this requirement by using two different portions of 
the same image.

3) Field counts should be computed in a region far enough from
the clusters to not be contaminated by cluster galaxies, but near enough to take
into account nonuniformities of the field on the scale of
the angular size of the cluster.

   \begin{figure*}[]
   \resizebox{\hsize}{!}{\includegraphics{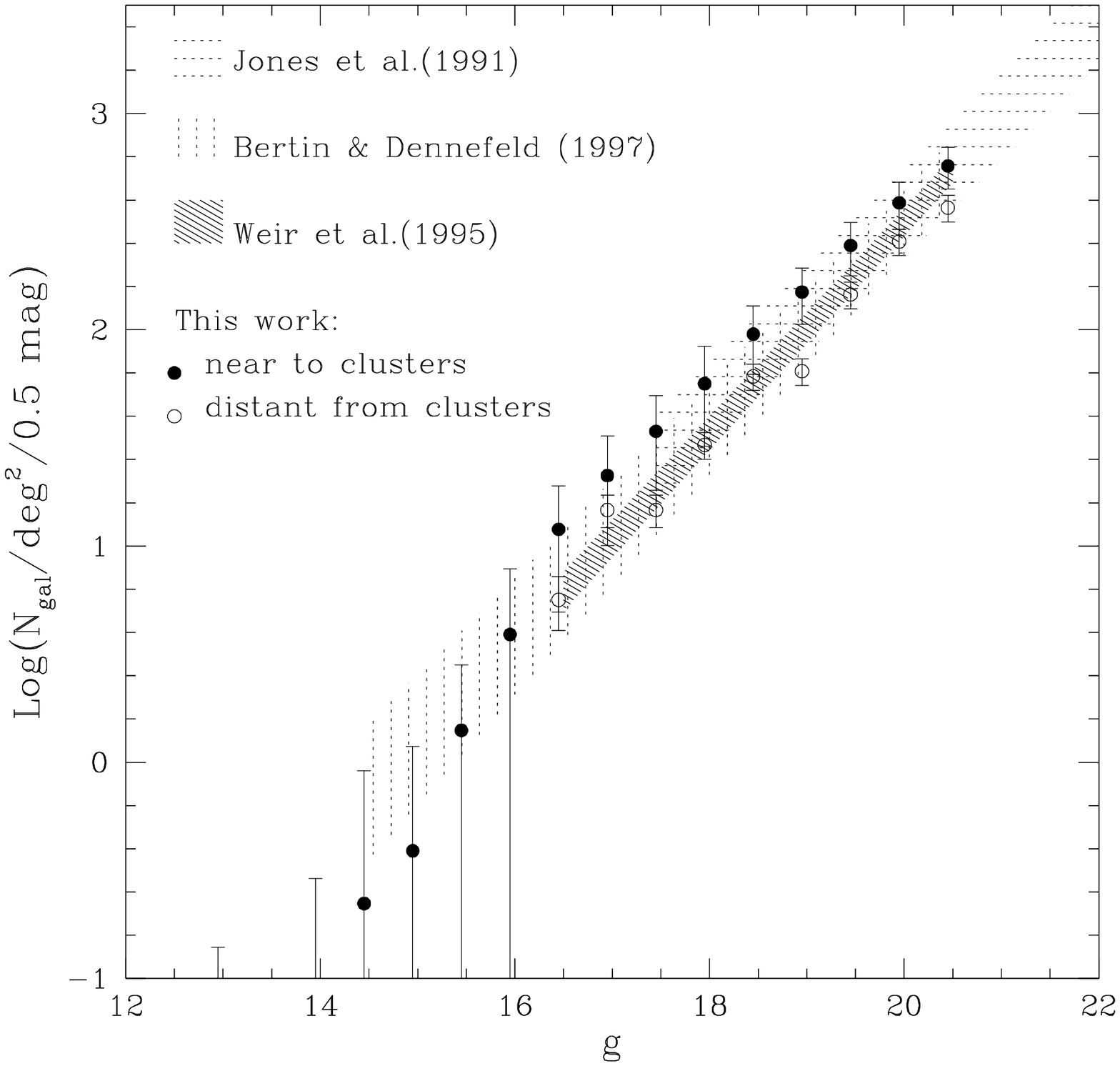}
   \includegraphics{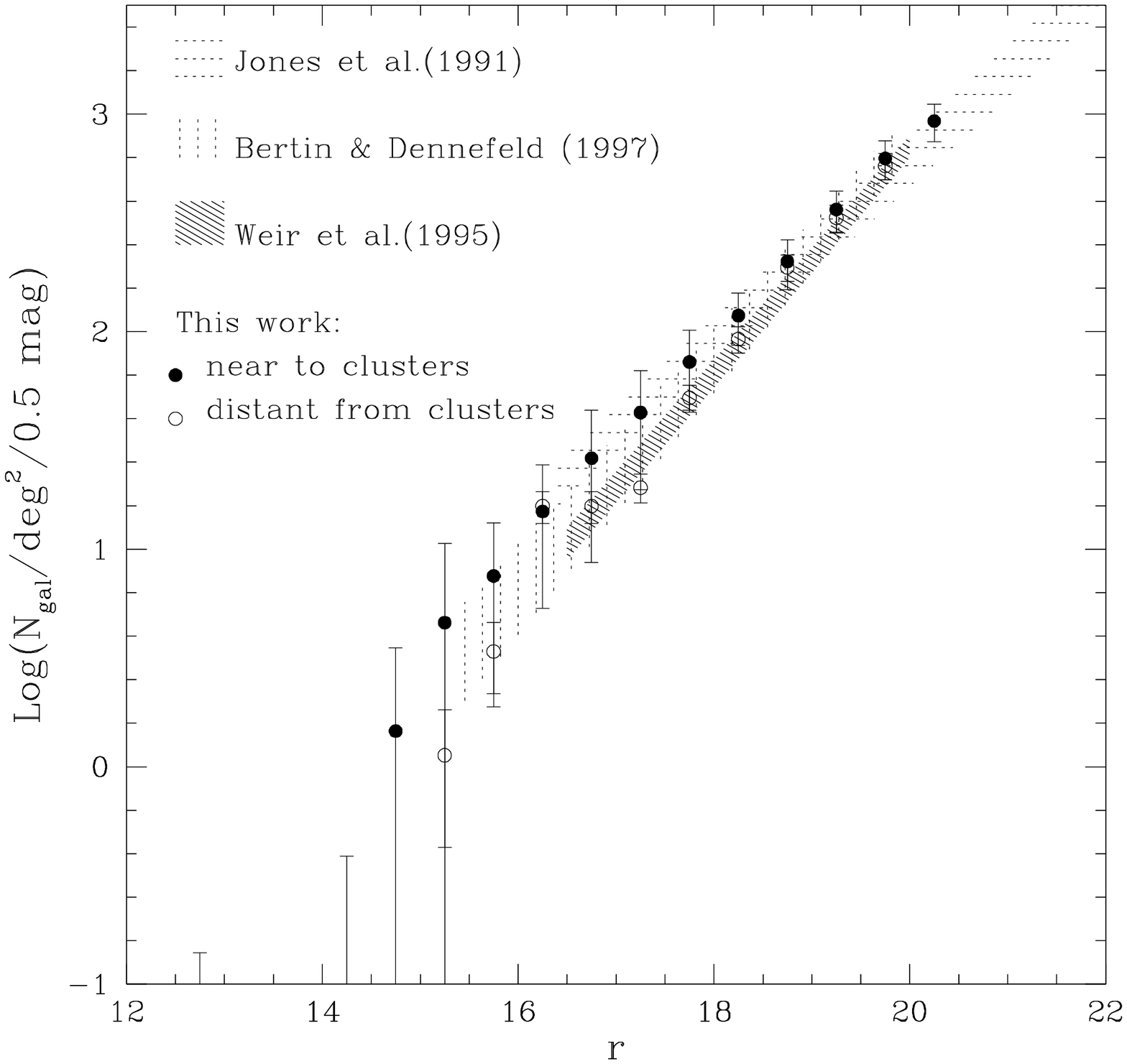}}
      \caption[]{Mean background counts in regions near and far
	from clusters, compared to literature, in the $g$ and $r$ band (see discussion in
	text). Literature counts fall within the shaded regions.}
         \label{Fbkgcounts}    \end{figure*}

For this purpose we take advantage of our wide field capabilities by dividing the 
$10\times 10$ Mpc region in two parts:  an
inner circle of 3 Mpc radius, i.e. $1$ Abell radius ($R_A=1.5 h^{-1}$ Mpc), 
used to search for the cluster overdensity, and an external region, beyond the 3 Mpc circle
and within the 10 Mpc square, where we measure the field counts (Fig.\ref{a152}).
This radius is large enough that
the contamination of cluster galaxies in the control field line of sight is
minimal and the S/N of the LF is not significantly reduced (if at all).
In fact, the cluster overdensity is undetected at these radii for the
large majority of the clusters.

However, a control field too near the cluster, while diminishing
the S/N of the LF determination, does not alter the shape of the LF. In fact:
\begin{eqnarray}
\label{eq1}
N_{cl}(m) & = & N_{cl+bkg}(m)-N_{bkg}(m)=\nonumber\\
& = & (N_{cl}(m)+N_{bkg}(m))-N_{bkg}(m)
\end{eqnarray}
where $N_{cl}(m)$ is the number of cluster galaxies at a certain magnitude $m$ and $N_{bkg}(m)$ is the number of background galaxies at the same magnitude.
A too near control field is contaminated by the cluster galaxies, i.e.
\begin{equation}N_{bkg}^{near}(m)=N_{bkg}(m)+\gamma\cdot N_{cl}(m)\end{equation}
where $\gamma$ is the ratio between the cluster galaxy density in the cluster
region and in the control field direction.
Therefore using $N_{bkg}^{near}(m)$ instead of $N_{bkg}(m)$, 
one obtains from eq.\ref{eq1}: 
\begin{eqnarray}
N_{cl}^{near}(m) & = & N_{cl+bkg}(m)-(N_{bkg}(m)+\gamma\cdot N_{cl}(m))=\nonumber\\
& = & N_{cl}(m)\cdot (1-\gamma)
\end{eqnarray}
i.e., by using a too near control field, the number of cluster galaxies (and
thus the LF if the applied k-correction does not depend on galaxy type, as in our case) is simply diminished by a multiplicative factor
and its shape is not affected. To be more precise, one should note that
we are supposing that the LF shape does not depend on the
cluster location or that the contamination is very small, so that
even large differences in the LF have null impact.

Field galaxy counts are measured in the external region
once the removed areas are taken into account. 
Furthermore, in order not to bias the average background due to the
existence of other groups and clusters, we remove every density peak above 
$2\sigma$ from the mean field density (see Fig.\ref{a152}). The average, which we call 
the ``local field'', is the adopted estimate of the background counts in the cluster
direction.

The ``local'' field, computed all around the cluster, is a better measure of the
contribution of background galaxies to counts in the cluster direction than the
usual ``average" field (measured on a single spot and/or far from the
considered cluster), since it allows us to correct for the
presence of possible underlying large--scale structures both at the cluster distance
and in front of or behind it. 

Fig.~\ref{Fbkgcounts} shows that our galaxy counts are
consistent with previous determinations, and in
particular with those by Weir, Djorgovski \& Fayyad (\cite{Weir95b}), 
who also made use of DPOSS plates.
Nevertheless, counts near clusters (filled
circles), but not too near to be affected by them, tend to be systematically higher 
than the average and in particular of those extracted in a reference
region particularly devoid of structures (empty circles), even if differences
are
within the errors. This difference can be as high as 80\% of the mean value.
The higher value can be explained by the fact that we are sampling the superclusters surrounding
the studied clusters, whose contribution can be missed when measuring background in smaller and random fields, 
as often done in the literature.

Once the background to be subtracted from cluster counts has been determined, we
need a robust evaluation of the error involved in the subtraction process.
There are three sources of errors:
Poissonian errors for galaxies belonging to the cluster, plus
Poissonian and non--Poissonian fluctuations of the background
counts.

Poissonian fluctuations in the number of cluster members are significant only at
magnitudes where the control field counts have close to zero
galaxies per bin. Poissonian fluctuations of the background in the control field
direction are small because of the large area used to determine the local field
(at least 20 times larger). Therefore, the dominant term in the error budget 
is due to the non--Poissonian fluctuations of background counts
along the cluster line of sight. The wide coverage of
the DPOSS fields allow us to easily and directly measure the {\it variance} of galaxy counts,
and thus the field fluctuations (Poissonian and non-Poissonian) on the angular scale of each
individual cluster in adjacent directions, instead of relying on model estimates 
(e.g. Huang et al. \cite{Huang97}). 
It should be noted that, until a few years ago, non-Poissonian fluctuations were
often completely ignored, thus underestimating the errors on the LF.

Adami et al.(\cite{Ada98}) questioned this statistical method of computing the
LF (a method that dates back at least to Zwicky \cite{Zwi57}), and checked the validity of a
statistical field subtraction by means of a redshift survey in the case of one single
cluster, finding a discrepancy between the counts of
Bernstein et al.(\cite{Ber95}) and those inferred from spectroscopic measurements.
They argued that the statistical method can be affected by potential errors.
Nevertheless their narrow field of view ($7.5^{\prime}\times
7.5^{\prime}$, $\sim 280\times 280$ Kpc$^2$ at cluster distance) and the small number of
galaxies (49) in their sample, does not allow to draw any significant conclusion
from this test (the observed discrepancy is statistically significant only
at the $1\sigma$ level).

Furthermore, the fact that they sampled the core of a rich cluster where other effects
--as they notice-- such as tidal disruption might be dominant, and the possibility that the
Bernstein et al. field counts could be underestimated (the use of random fields to measure 
background does not take into account the presence of underlying large-scale 
structure), contribute to bringing the discrepancy well below $1\sigma$.

   \begin{figure*}[t]
   \centering
   \includegraphics[width=17cm]{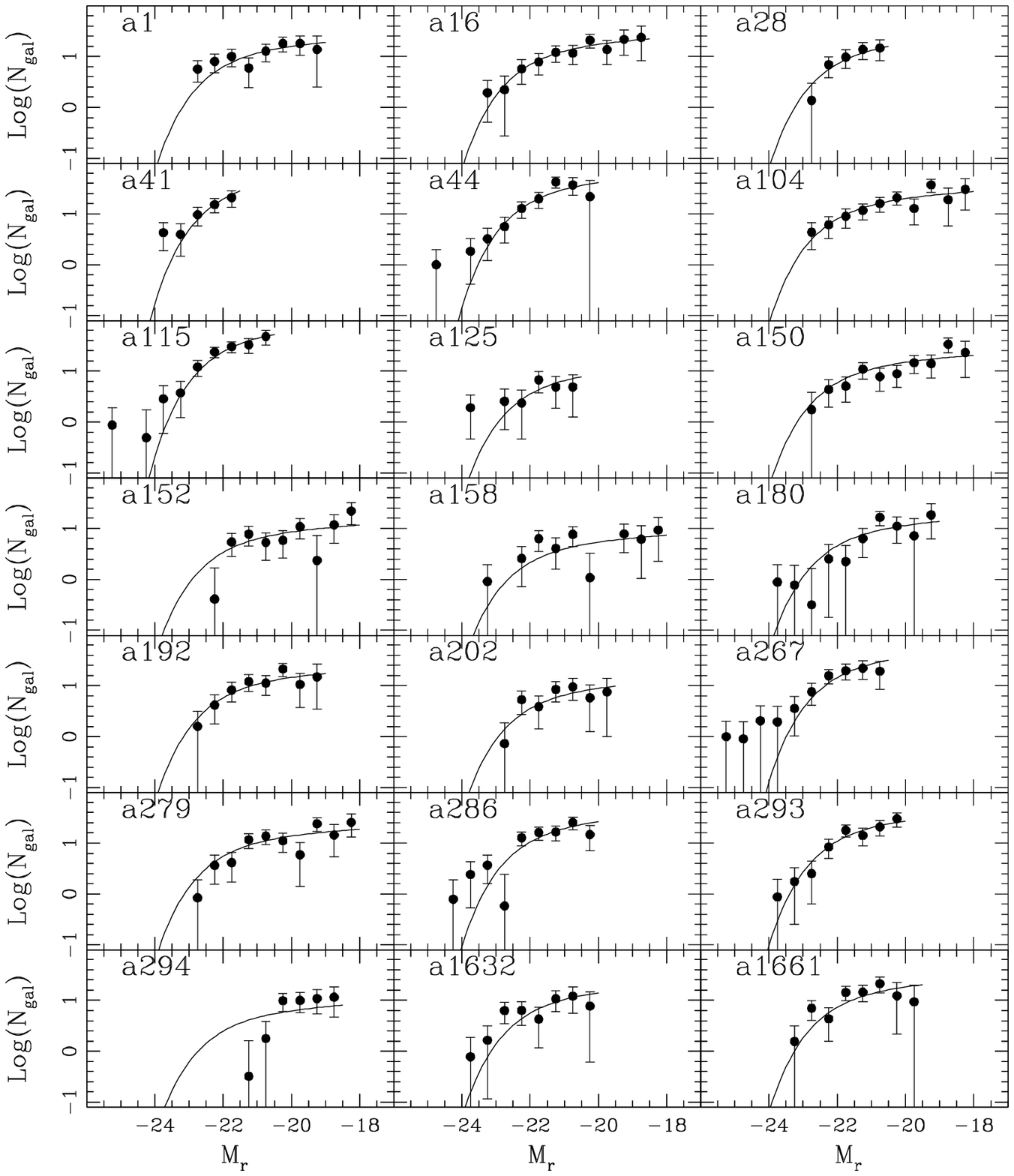}
      \caption{{\bf (a)} The background-corrected galaxy counts for the first
      21 cluster
      of our sample in the $r$ band. The best-fit Schechter function 
      of the cumulative LF (par.4), normalized to the total counts in each cluster,
       is shown as a continuous line.}
         \label{lf1}    \end{figure*}

\subsection{The individual LF}
Due to the low (1 arcmin) astrometric
accuracy of the Abell (\cite{Abell58}) cluster 
centers, we first search for the central $1.5 \sigma$ density peak in the
inner 3 Mpc circle (much larger than 1 arcmin at all considered redshifts) centered
on the approximate cluster position as determined by Abell
(Fig.\ref{a152}). We then derive the cluster LF by subtracting, 
from the galaxy counts measured in this region, the local field counts,
rescaled to the effective\footnote{I.e. the area corrected for the regions removed
around the troublemakers.} cluster area.

This approach allows us to take into account the cluster morphology
without having to adopt a fixed cluster radius, and thus to apply the local field correction to the
region where the signal (due to the cluster) to noise (due to field and
cluster fluctuations) ratio is higher, in order to minimize statistical uncertainties.

The LF for individual clusters in the $r$ band are shown in Fig.\ref{lf1}a,b,
together with the best-fit Schechter function of the {\it composite} LF
(sec.\ref{results}). 
Because we already used the whole cluster for computing the LF, 
individual LFs cannot
be improved further, except by performing expensive redshift surveys.

   \begin{figure*}[t]
   \resizebox{\hsize}{!}{\includegraphics{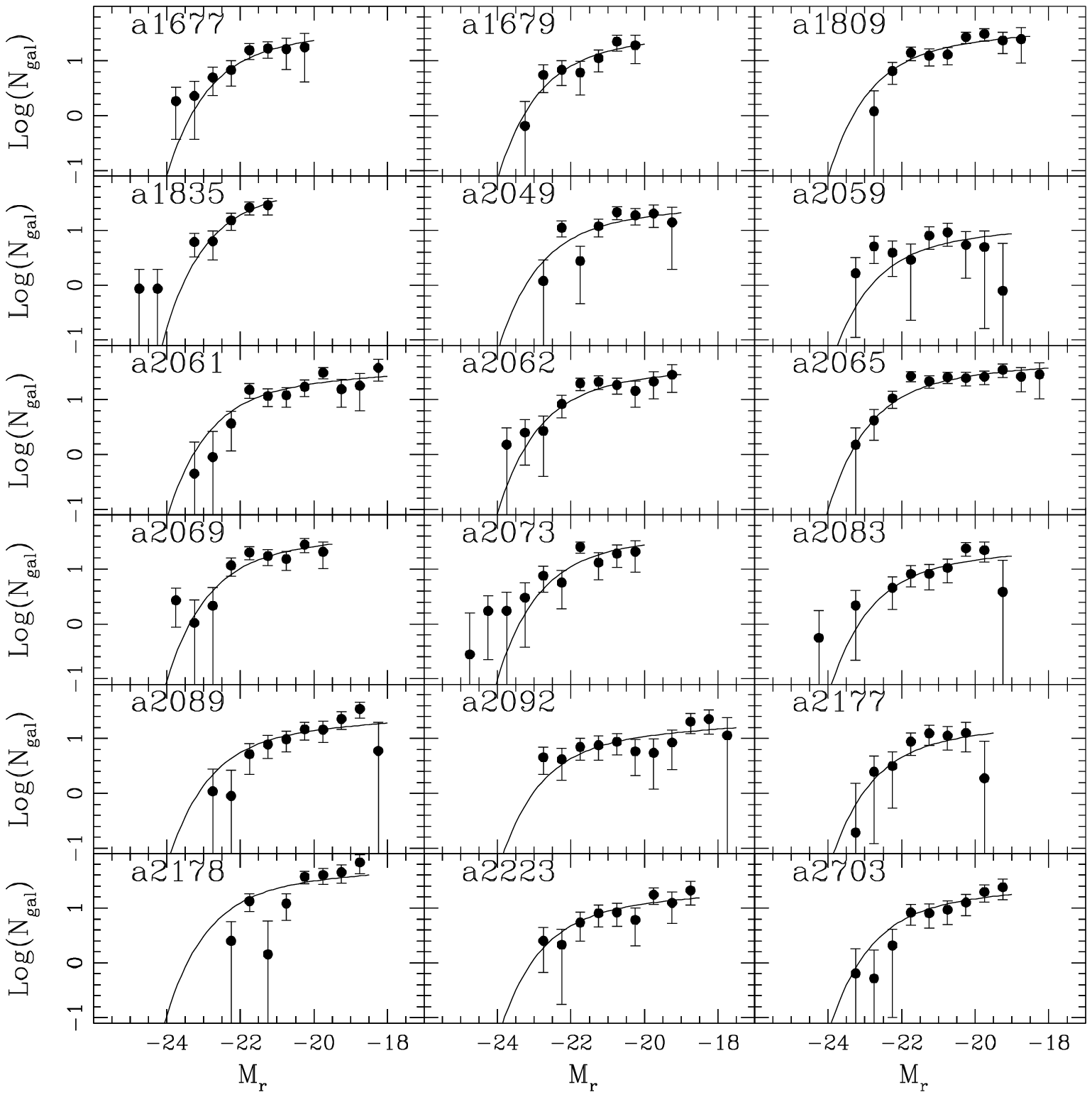}}
      \addtocounter{figure}{-1}
      \caption[]{{\bf (b)} As in Fig.\ref{lf1}a for the remaining 18 clusters.}
	\end{figure*}

   \begin{figure*}[!t]
   \resizebox{\hsize}{!}{\includegraphics{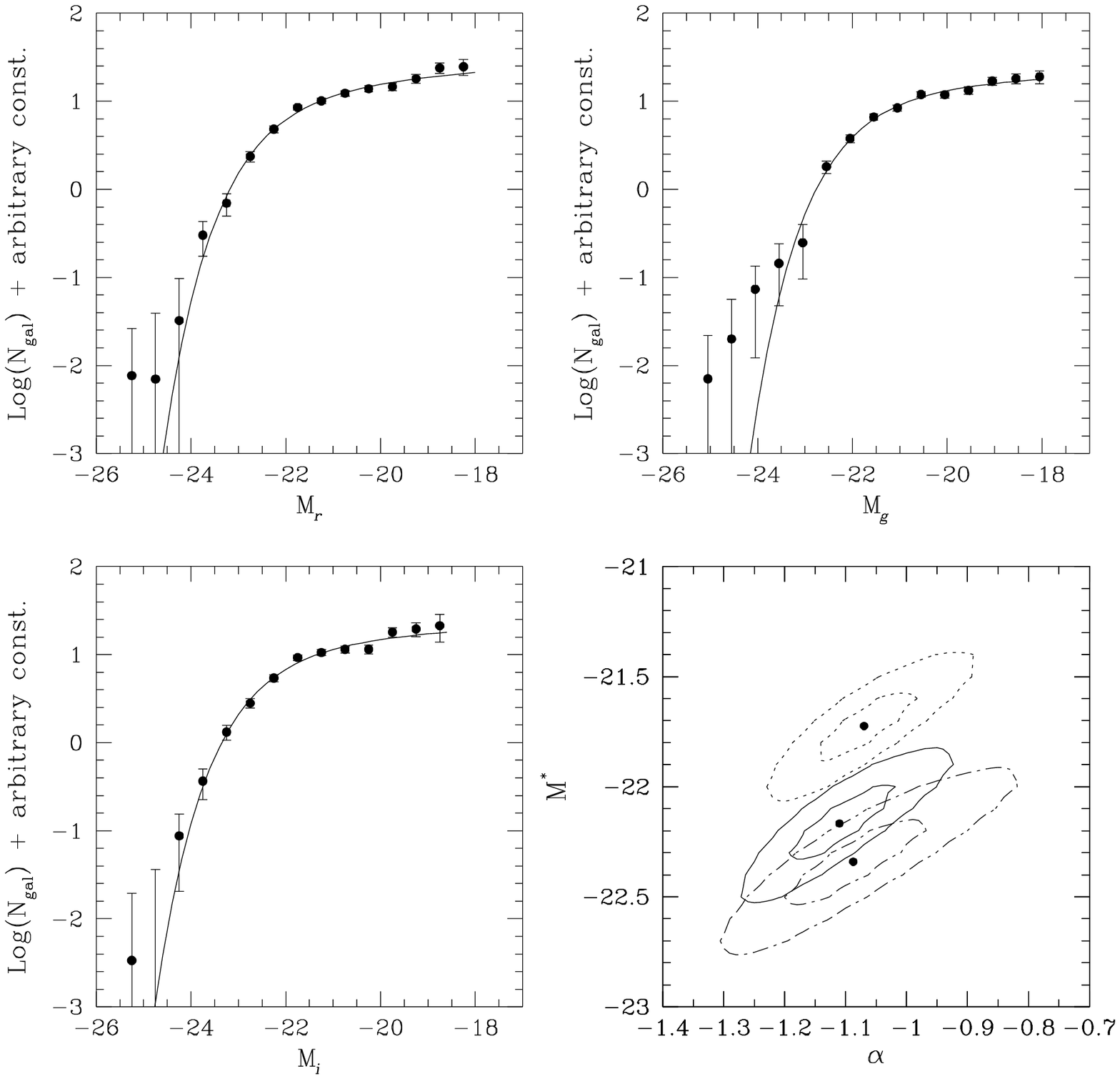}}
      \caption[]{The composite LF in the $g$, $r$ and $i$ bands obtained
	excluding the brightest member of each cluster (filled
	dots). The best fit Schechter functions  are represented by
	continuous lines, with the 68\% and 99\% confidence levels
	of the best fit parameters in 
	the bottom right panel ($g$: dotted line; $r$: continuous line; $i$:
	dashed-dotted line).}
         \label{cum_tot39}    \end{figure*}

\section{Results}
\label{results}

Most of our clusters have too few galaxies to accurately
determine the shape of the LF. Instead, we can combine all  individual LFs to
construct the composite LF of the whole sample. In doing so, vagaries of
individual LFs are washed out and only the underlying possibly universal LF is
enhanced. We adopt the method used by GMA99, consisting
of a modified version of the formula introduced by Colless (=C89, \cite{Coll89}).
In practice, the composite
LF is obtained by weighting each cluster against the relative number of galaxies in a
magnitude range that takes into account the variations in
the completeness limit of our data. 

Ostriker \& Hausman (\cite{Ostr77}) have shown that giant galaxies in clusters may be
the result of peculiar accretion processes. For this reason we took care to remove
from each cluster the Brightest Cluster Member (BCM).

The final composite LF is shown in Fig.~\ref{cum_tot39} for the $g$, $r$
and $i$ bands.

The fit of the composite luminosity functions to a Schechter (\cite{Sche76}) 
function\footnote{The function has been convolved with a boxcar 
filter in order to take into account the finite amplitude of the magnitude bins.} 

$$N(M)=\Phi\times 10^{0.4(M^*-M)(\alpha+1)}\exp(10^{0.4(M^*-M)})$$
gives the values listed in Table \ref{partab}, where $M^*$ is the characteristic
knee magnitude and $\alpha$ is the slope of the LF at faint magnitudes. Fig.
\ref{cum_tot39} shows the three best-fit functions together with the  $68\%$ and
$90\%$ confidence levels. The LFs turn out to be quite well described by a
Schechter function in our magnitude range (see $\chi^2$ in Tab.\ref{partab}).

   \begin{figure*}[th]
   \resizebox{\hsize}{!}{\includegraphics{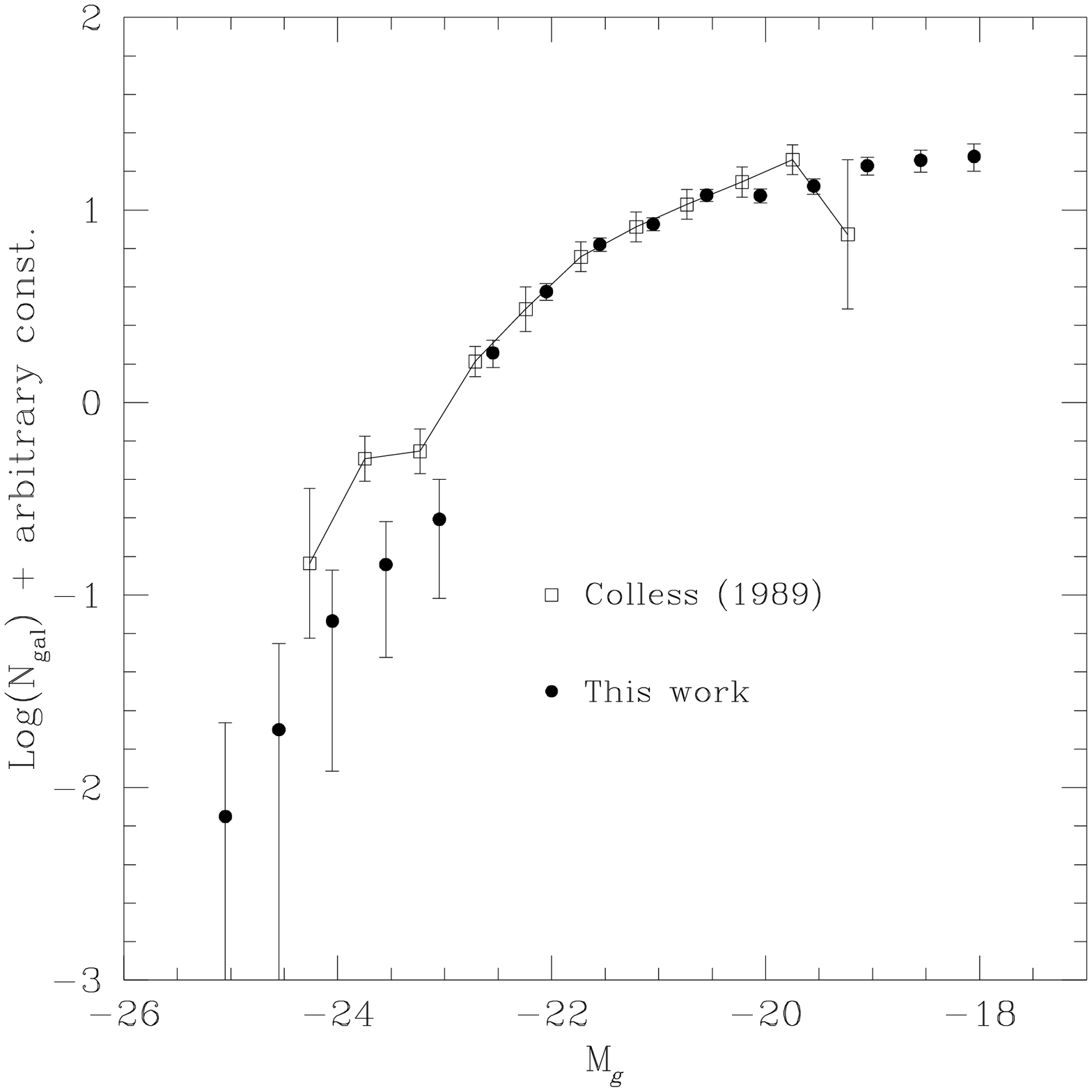}
   \includegraphics{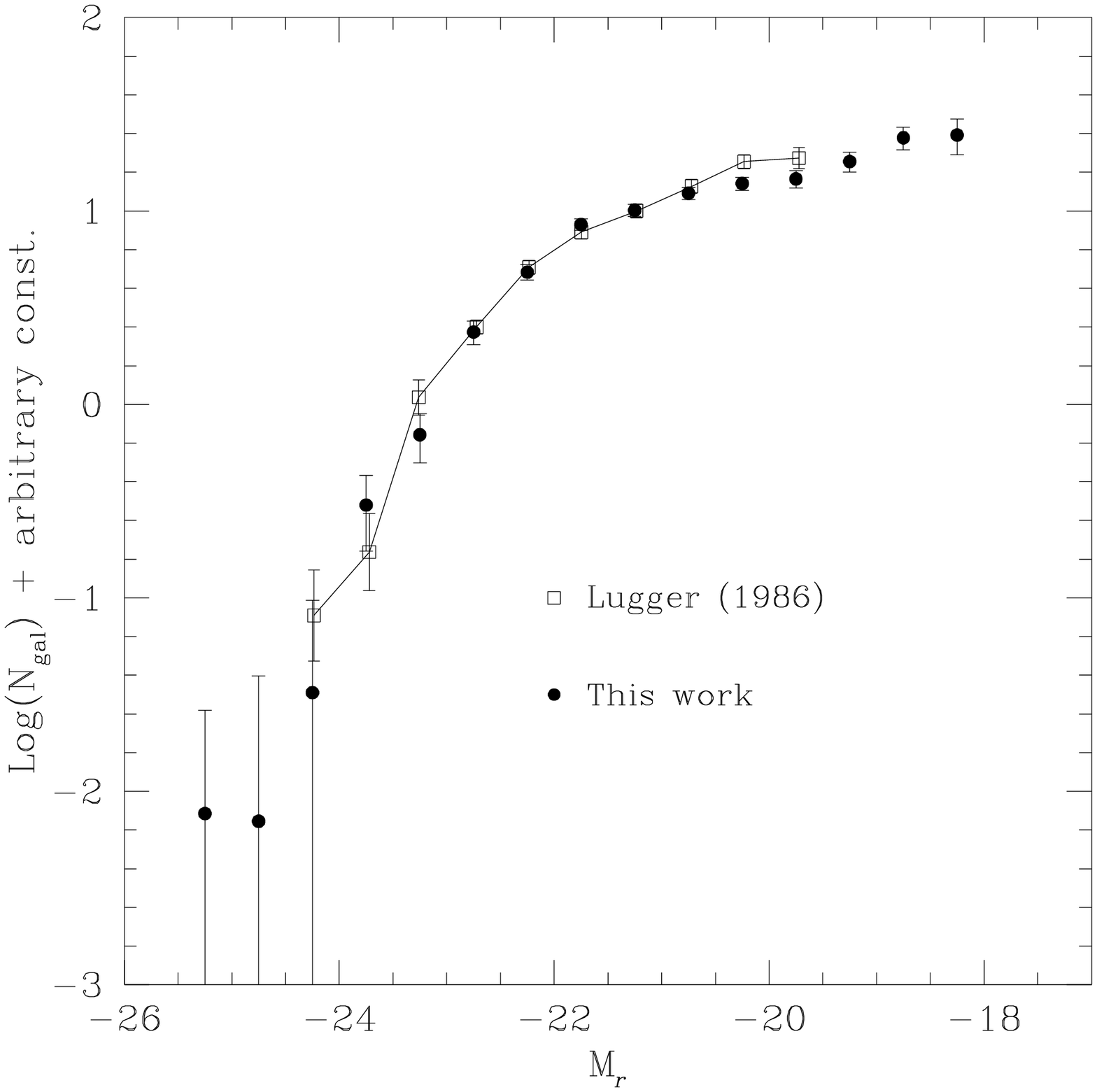}}
      \caption[]{Comparison of our composite LF with those of Colless (1989) in
	the $g$ band and Lugger (1986) in the $r$ band, both based on
	photographic data. Literature LF have been vertically shifted to
	match our LF.}
         \label{colless}    \end{figure*}

   \begin{table}
   \caption[]{LF best fit Schechter parameters. The given errors are
		referred to 1$\sigma$ confidence levels.}
   \label{partab}
   \begin{tabular}{cccc}
\hline
\\
{Band} & {$\alpha$} & {$M^*$} & {$\chi^2$/d.o.f.}\\
\\
\hline
\\
$g$ & -1.07$^{+0.09}_{-0.07}$ & -21.72$^{+0.13}_{-0.17}$ & 9.4/13\\
$r$ & -1.11$^{+0.09}_{-0.07}$ & -22.17$\pm$0.16 & 10.2/13\\
$i$ & -1.09$^{+0.12}_{-0.11}$ & -22.35$\pm$0.20 & 11.4/12\\
\\
\hline
	\end{tabular}
\end{table}

The faint end of the composite LF ($g$: $-1.07^{+0.09}_{-0.07}$, $r$: $-1.11^{+0.09}_{-0.07}$, $i$: $-1.09^{+0.12}_{-0.11}$) is, in all bands, shallower than the
traditional value, $\alpha = -1.25$ (cf. Schechter \cite{Sche76}),
but still compatible within the 99\% level in the $r$ and $i$ bands.
The best fit values of $\alpha$ in the three bands are almost identical, while $M^*$
increases from the blue to the red band as it is expected from the color of the dominant
population in clusters (taken, for example, from Fukugita et al. \cite{Fuk95}).

In order to test if our background is measured too near the
cluster, we re--computed the LF by adopting the $g$ and $r$ field counts derived
by Weir et al (\cite{Weir95b}), from the same photographic
material and using the same software. We also adopt our
direct measure of the background fluctuations, because these
are not provided in Weir et al. The newly found best fit
parameters differ by less than $1\sigma$ from those previously determined, 
thus suggesting that cluster members that are more than 3 Mpc away from
the clusters (and that therefore fall in our control field direction) have
null impact on the composite LF. A definitive assessment of the effects
of this assumption on the outer LF, 
which is much more sensitive to a small error on the background correction,
calls, however, for a larger sample of clusters.
   \begin{table}[htb]
   \caption[]{The best-fit Schechter parameters for the
    mean-background corrected LF.}
   \label{meantab}
   \begin{tabular}{cccc}
\hline
\\
{Band} & {$\alpha$} & {$M^*$} & {$\chi^2$/d.o.f.}\\
\\
\hline
\\
$g$ & -1.11$\pm$0.07 & -21.87$\pm$0.13 & 9.7/13\\
$r$ & -1.12$\pm$0.06 & -22.20$\pm$0.13 & 11.2/13\\
\\
\hline
	\end{tabular}
\end{table}

We stress that for the time being we prefer to use the local background 
for an aesthetic reason: the use of the far
field implicitly assumes that all galaxy overdensities near the
cluster belong to the cluster, including superclusters and filaments.
From a technical point of view, the problem is similar to the well
understood problem of performing accurate photometry of non isolated
objects: when an object in embedded in (or simply superposed to)
a much larger one as it happens, for instance, in the case of HII
regions or globular clusters on a galaxy or in that of a small galaxy projected on the halo of a larger one. It makes no sense to measure the background very far from the source of interest, since such a procedure ignores the non negligible background contributor. By using a
``far distant" background field, we would produce perfectly empty regions at the location
of clusters in superclusters, for HII regions in galaxies, and at every locations in
the Universe where there are superposed structures of different sizes.

\subsection{Comparison with previous determinations}

Our composite $g$ and $r$ LFs can be easily compared with those obtained from
photographic material by C89 in the $B_J$ band and by
Lugger (=L86, \cite{Lugger86}) in the $R$ band, as shown in Fig.\ref{colless}.
Conversions between their photometric systems
and our own has been performed using the color conversions given in the original
papers and those by Fukugita et al.(\cite{Fuk95}).
We found that the characteristic magnitudes
agree very well (within 1$\sigma$) while the faint end slopes
are compatible within 2$\sigma$ ($\alpha_{C89}=-1.21$ and
$\alpha_{L86}=-1.24$). At bright magnitudes our LF matches 
the Lugger one well, but not the Colless one, which includes the BCMs in the LF.
Anyway, our LF extends more than one magnitude further both at the
bright and faint end: the bright end, which includes rare objects, is better
sampled due to the large area coverage of our survey, whereas fainter
magnitudes are reached due to our deeper magnitude limit.

\begin{figure}[t]
   \resizebox{\hsize}{!}{\includegraphics{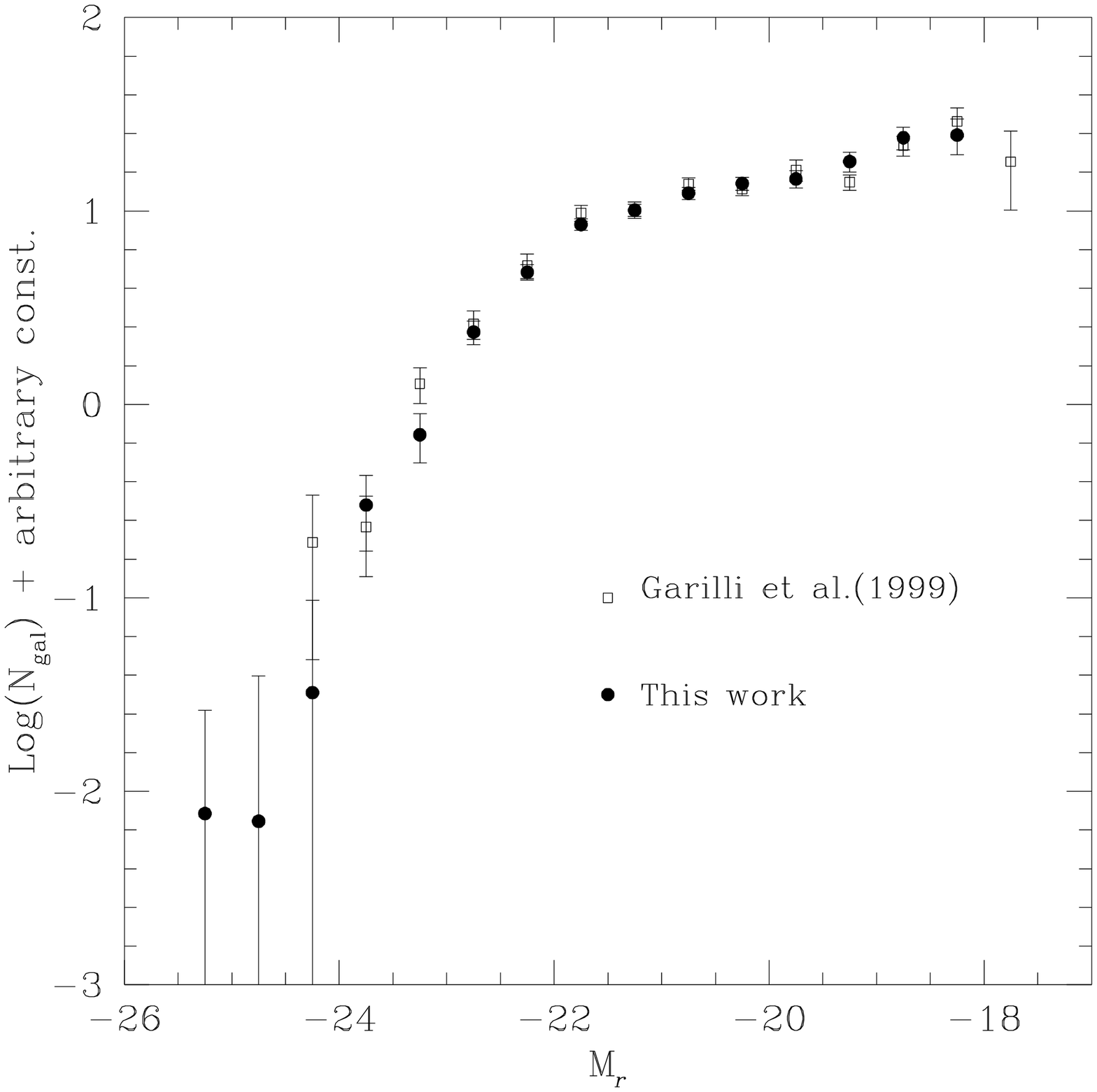}}
      \caption[]{Comparison between our composite LF and
	the Garilli et al.(\cite{Garilli99}) LF
	obtained from CCD data and adopting a different method to remove
	interlopers (see discussion in text).}
         \label{ccd_conf}    \end{figure}

Evidence in favor of a flat LF has been presented by many authors (cf. Gaidos
\cite{Gai97} for 20 Abell clusters and GMA99 for 65 clusters).
A comparison with GMA99 is of particular interest since, in addition to 
adopting our same photometric system, they use
a completely different method for removing possible interlopers. GMA99 exploit the fact that
the observed colors of the galaxies change with redshift due to the K correction,
which moves the background objects in a locus of the color--color plane 
different from that occupied by the cluster galaxies.
We compare our $r$ band LF with GMA99 in Fig. \ref{ccd_conf}.
The agreement is impressive considering not only that the background correction
is made using different approaches, but also the different total magnitude
corrections (FOCAS ``total" in this work, aperture magnitude corrected to total in
GMA99). Moreover, our sample is independent from theirs, except for a few
clusters wich are anyway sampled in different
regions due to the different field of view.

We find that both the slope and the characteristic magnitude of their best-fit function 
are in good agreement with ours and are compatible within the errors (within
$2\sigma$).
This agreement tends to confirm that our choice of using a local
background determination instead of the ``average'' one leads to a good  estimate of the number of interlopers contaminating cluster galaxy counts.

In the comparison with GMA99, a few more facts are worth mentioning:\\
 -- we reach a similar determination of the cluster LF at lower telescope time price:
using just a few ($\sim10$) plates taken for  a very general purpose (a sky survey) with
a Schmidt telescope, we achieve the same performances as in a multi--year CCD
campaign on a 2m class telescope;\\
-- our LF extend to brighter magnitudes, thus sampling  the LF
at the location of rare objects, a possibility allowed only by large area surveys.\\

Even though CCD  data are usually deeper
and have higher photometric accuracy than ours,  they are also limited to small
regions of the clusters and usually cover different cluster portions at different
redshifts.  Our data, instead, cover the whole cluster area independently of the
redshift, but are selected in apparent magnitude.  This means that the fainter
magnitude bins of the composite LF, are populated mainly by the galaxies in the
nearer clusters.
In absolute--magnitude selected surveys instead, the faintest bins usually include
preferentially distant galaxies, due to the large area covered at high redshift
with a fixed field of view. This is not true for GMA99, where the redshift distribution
was quite uniform since the authors tried to observe 
nearby clusters with a large field of view and
distant clusters are slightly less deeply probed than the near ones.\\
However, the total number of objects in our sample is approximately 1.5 
times the number of objects in the GMA99 sample.

Our results disagrees with the steep ($ -1.6 <\alpha <
-1.4$) LF found by Valotto et al.(\cite{Val97}). Their work
is based on photographic data taken from the APM cluster survey 
and they adopt, as we do, ``local'' background counts
measured in annuli surrounding each cluster. 
Nevertheless, their completeness limit is 1.5 magnitudes shallower than ours,
so that they are sampling the brighter portion of the LF, and therefore the slope
is subject to large errors.

At first glance, our claim that a Schechter function is a good fit to
our data ($\chi^2 \leq 1$) seems in contradiction with 
various claims of a non--universal LF produced by the various morphological
composition of clusters and by the non--homology of the LFs of the morphological types
(e.g. Sandage, Bingelli \& Tammann \cite{SBT85}; Jerjen
\& Tammann \cite{Jer97}; Andreon \cite{And98}) or because of the variable dwarf content of clusters
(Secker \& Harris \cite{Seck96}; Trentham \cite{Trent97}, \cite{Trent98}).
   \begin{figure}[t]
   \resizebox{\hsize}{!}{\includegraphics{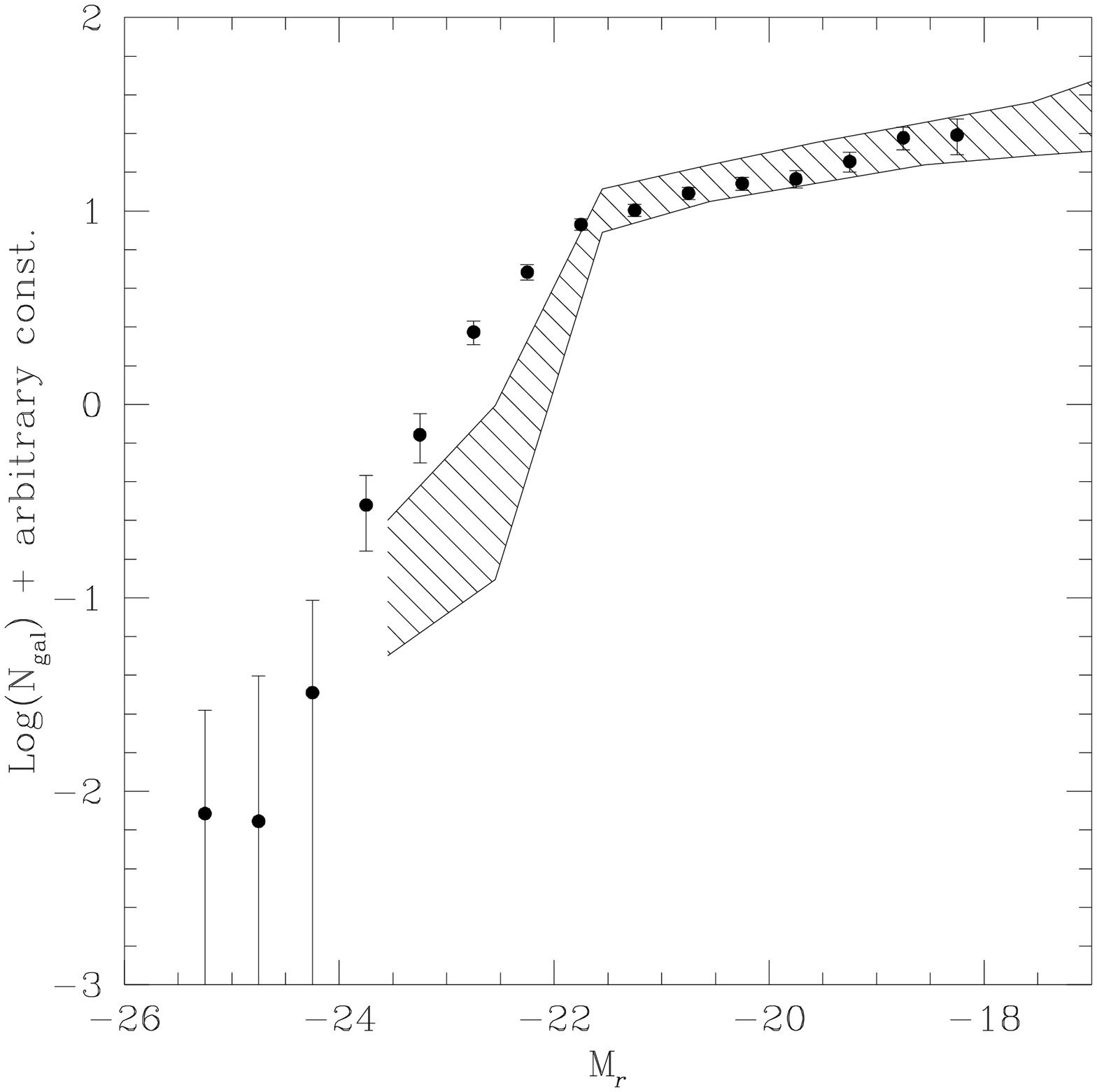}}
      \caption[]{Comparison between our composite LF (filled dots) and
	the Trentham (\cite{Trent97}) LF (shaded region) based on CCD data.}
         \label{trent_conf}    \end{figure}

   \begin{table*}[]
   \caption[]{Results of $\chi^2$ tests for different subsamples.}
   \label{chi}
   \begin{tabular}{c|cc|cc|cc}
\hline
& \multicolumn{6}{c}{Band}\\
 & \multicolumn{2}{c}{$g$} & \multicolumn{2}{c}{$r$} &
\multicolumn{2}{c}{$i$}\\
\hline
 & {$\chi^2$/d.o.f.} & {Prob.$>\chi^2$} & {$\chi^2$/d.o.f.} & {Prob.$>\chi^2$} &
{$\chi^2$/d.o.f.} & {Prob.$>\chi^2$}\\
\hline
\\
$R>1$ vs $R \leq 1$ & 22.98/16 & 0.11 & 6.58/15 & 0.97 & 11.38/14 & 0.66\\
BM I+I-II vs II+II-III & 21.91/15 & 0.11 & 13.18/12 & 0.36 & 11.43/11 & 0.41\\
BM I+I-II vs III & 22.66/14 & 0.07 & 20.81/12 & 0.05 & 13.29/11 & 0.27\\
BM II+II-III vs III & 9.45/15 & 0.85 & 9.37/14 & 0.80 & 4.18/12 & 0.98\\
Compact vs Elongated & 20.47/14 & 0.12 & 11.78/14 & 0.62 & 14.59/13 & 0.33\\
Compact vs Multiple & 17.74/17 & 0.41 & 6.92/15 & 0.96 & 17.97/14 & 0.21\\
Elongated vs Multiple & 20.03/14 & 0.12 & 11.49/14 & 0.65 & 10.13/12 & 0.60\\
\\
\hline
	\end{tabular}
\end{table*}
Trentham (\cite{Trent97}), for instance, showed that the cluster LF rises steeply at
faint magnitudes ($M_g>-18$) and thus a simple Schechter function cannot 
properly describe the whole distribution. 
Nevertheless, as shown in Fig. \ref{trent_conf}, for magnitudes between -22  and -17  the LF is
quite flat and in good agreement with our data.
In fact, in our magnitude range the contribution of dwarf galaxies is visible
only in the faintest bins, as suggested by the fact that in Fig. \ref{cum_tot39} the
last points lie systematically above the best-fit function. This trend (a flattening of
the distribution around $M=-21$ and a steepening over $M=-19.5$) is also confirmed 
by the comparison with GMA99, whose LF shows a similar behavior.
At bright magnitudes, instead, the act of averaging over the cluster region can mask the 
environmental effects.

We must note that while in Fig.\ref{trent_conf} the two LFs differ substantially at 
the bright end, our data are in very good agreement with L86 and GMA99, thus suggesting
that Trentham is underestimating the contribution of bright galaxies to the LF. This can be due
to various reasons, including the small area and the specific portions of the clusters sampled, or 
the different morphological composition of his clusters.
Moreover, due to our larger number of clusters, we can sample the LF at twice
the resolution in magnitude.

\subsection{Dependence on the cluster physical parameters}

We compared the LFs obtained dividing our sample into rich (Abell class
$R > 1$) and poor ($R \leq 1$) clusters. Tab. \ref{chi} shows that
the LFs of these two classes are consistent within the
errors. GMA99 found instead that the slope of the LF computed in the 
central regions of the clusters depends on the cluster central
density, while they found mild differences, statistically
significant, between rich and poor clusters.
Our result differs from the GMA99 finding that 
the giant to dwarf ratio is higher in rich clusters than in poor ones,
but only in the statistical significance:
   \begin{figure}[]
   \resizebox{\hsize}{!}{\includegraphics{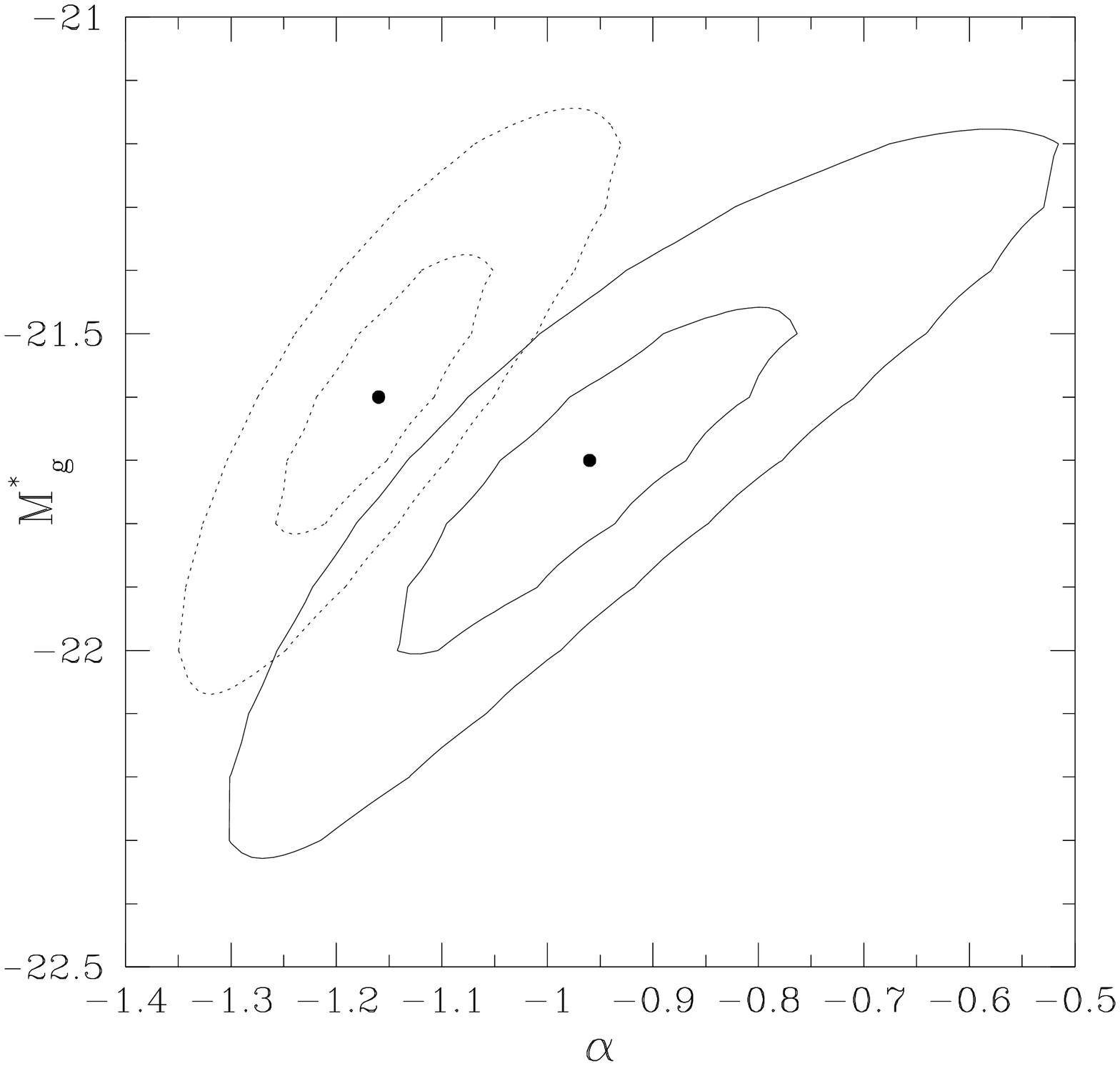}}
      \caption[]{The 68\% and 99\% confidence levels relative to the fit
	of the rich ($R>1$, continuous line) and poor ($R \leq 1$, dotted line)
	subsamples in the $g$ band.}
         \label{poorrich_Jmap}    \end{figure}
we find that poor clusters have a faint--end slope steeper than
rich clusters ($\overline{\Delta\alpha}=0.16$) by 
a quantity that is compatible within $1\sigma$ to those
derived by GMA99 in their poor--rich comparison. 
The dependence of the slope on richness is
more evident in the $g$ band, as shown in Fig.\ref{poorrich_Jmap}. 

We also explored the influence of the cluster dynamical state,
as indicated by the Bautz-Morgan type (Bautz \& Morgan
\cite{Bautz}), on the LF. We divided the sample in three subsamples:
BM I + BM I-II, BM II + BM II-III and BM III in order to have a similar
number of clusters in each group. We find that early and late BM types
have LFs which are
compatible within 95\%, in agreement with GMA99 and Lugger (1986).

Moreover, we divided our sample into 3 morphological classes based on 
visual inspection of density profiles (cf. sec. 4.1). We classified
clusters into ``compact'', showing a single strong density peak within the
1.5$\sigma$ isodensity contour above background; ``elongated'' if the
cluster is irregularly spread across the field with a weak density peak, and
``multiple'' if it shows multiple peaks. Again, we find 
no significant differences between the LFs of these classes of clusters.

In interpreting this result we note that when our sample is divided in subsamples
the number of objects may not be large enough for a $\chi^2$ test to reveal differences in the distribution, 
as in the case of the poor--rich comparison, so that a conclusive statement
calls for a larger sample. 

\subsection{Comparison with the field LF}

As already shown in GMA99, we find that the cluster LF is compatible with the field
LF. This result does not rule out environmental influence on galaxy
formation and/or evolution, but rather indicates that either evidence for such
effects must be investigated at fainter magnitudes than those reached
by DPOSS data, or that the effect is smaller than what the data
allows us to detect.
 
   \begin{figure}[t]
   \resizebox{\hsize}{!}{\includegraphics{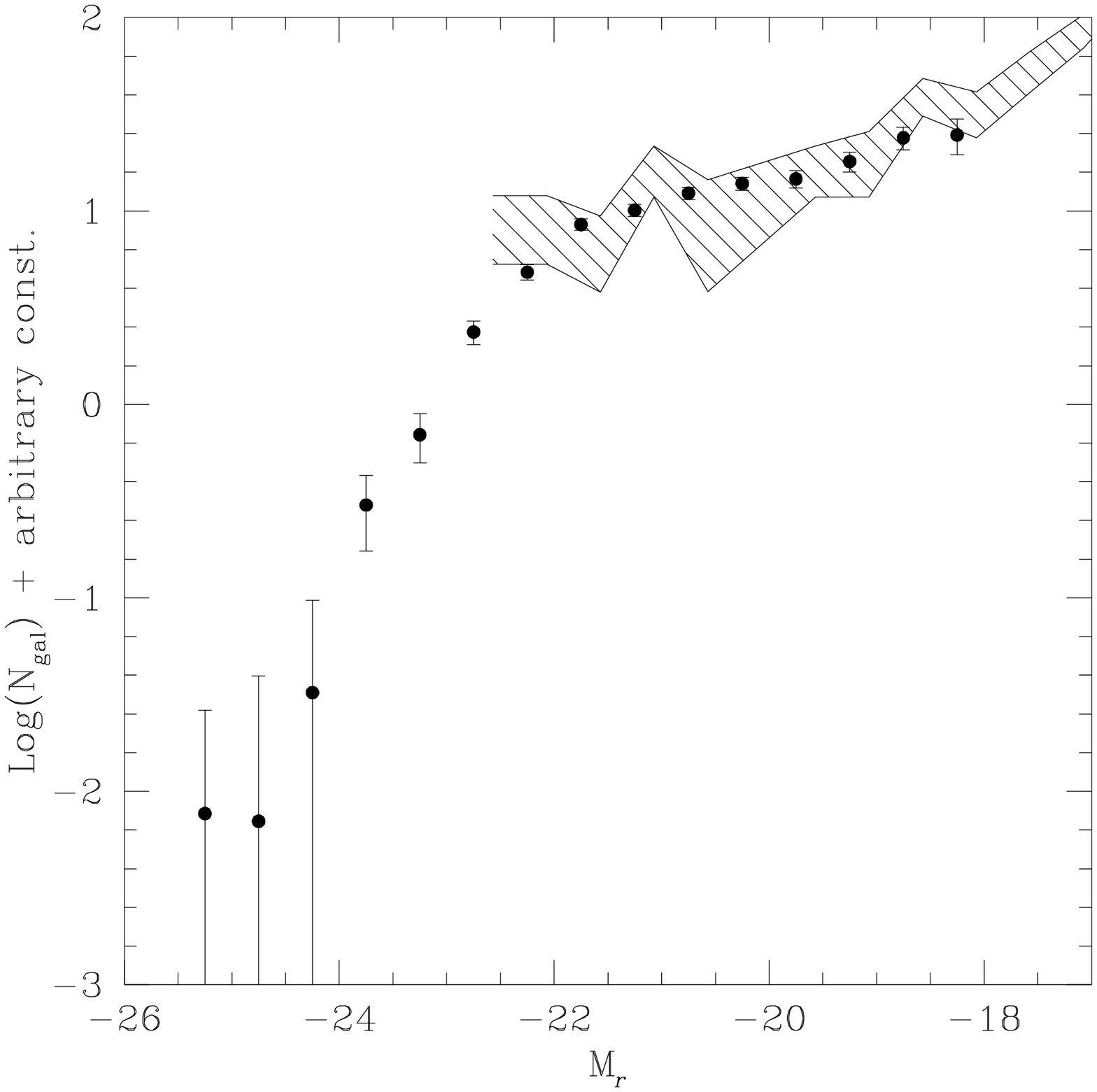}}
      \caption[]{Comparison between our composite LF (filled dots) and the LF obtained by 
      Andreon \& Cuillandre (\cite{And2000}), (shaded region).}
         \label{compact}    \end{figure}

\subsection{Compact galaxies misclassified?}

It could be argued that our (and also most literature) LF are flatter than
they should be since compact galaxies are misclassified as stars.
For the most distant cluster even normal galaxies are badly classified
due to the low angular resolution of the available images and/or to errors
in the star/galaxy classifiers (see, for example Drinkwater et al. \cite{Drink99}). In our case, the comparison with the LF of the
Coma cluster obtained by Andreon \& Cuillandre  (\cite{And2000})  settles this issue, because their determination is
not affected from this problem since it does not use any star/galaxy classification. Figure \ref{compact}
shows the good agreement between the two LFs and confirms that we are not missing any large population of compact galaxies.

\section{Conclusions}

We computed the composite LF of 39 clusters of galaxies at $0.08<z<0.3$ in three filters from
the DPOSS plates, using the well known fact that clusters are galaxy 
overdensities with respect to the field.
Our LF agrees with previous determinations of the cluster LF,
obtained using specifically tailored observations, while we use sky survey
plate
data. The LFs are well described by a Schechter function, with a shallow slope
$\alpha\sim-1.1$ with minor variations from blue to red filters and
$M^*\sim-22.4,-22.2,-21.7$ ($H_0=50$ km s$^{-1}$ Mpc$^{-1}$) in $g, r$ and $i$
filters, respectively.
The LFs are computed without the assumption of an average background
along the cluster line of sight, and use actual measurement of the background
fluctuation instead of relying on the formalism and hypothesis presented
in Huang et al. 1996 or, as in older works, assuming an 'average' error.
The existence of compact/misclassified galaxies have no impact on our
LF determination: they are a minority population or a magnitude
independent fraction of the number other galaxies.

The similarity of composite LFs by GMA99, measured from CCD photometry of the 
cluster central regions,
suggests minor differences between the LF in the cluster outskirts and in the central
one, or a minor contribution of galaxies in the cluster outskirts to the global LF. 

When our cluster sample is grouped in classes of richness, dynamical and
morphological type, we find no significant differences among the classes. However,
our cluster sample may be not large enough for detecting the differences found in
other studies, or the differences may be intrinsically too small to be detected
in a sample, like ours, which is large but not huge (and the latter sample still
does not exist). 

Our results on the cluster LFs are not completely new: other authors found
similar results, and for this reason we avoid repeating the cosmological
implication of our results. However, we wish to stress that:
we have a better control of the errors, due to the nearby control field
and the direct measure of the field variance;
we identify in the literature a few discrepant LFs in certain magnitude ranges;
we show that the statistical subtraction of the background is sound, since
we found the same LF shape found by Garilli et al., who removed interlopers
by adopting an independent method;
we obtain these results by using all-purpose photographic plates, instead of a multi-year CCD campaign.

We are currently increasing the present sample by an order of magnitude in order
to explore with greater statistical significance the dependence of the cluster
LF on the physical parameters.

\begin{acknowledgements}
Bianca Garilli, Dario Bottini and Dario Maccagni
are warmly thanked for providing us with the electronic access to the data
shown in Fig. 3 of Garilli et al. (\cite{Garilli96}). We also thank
M. Fukugita for providing us K-correction data and N. Trentham for the informations about the photometry in his articles.

The work on production and cataloguing
of DPOSS at Caltech was supported by a generous grant from The Norris
Foundation.  R. Gal acknowledges a partial support from a NASA Graduate
Fellowship.  We are also thankful to the POSS-II and DPOSS teams for their
efforts.
\end{acknowledgements}

\end{document}